\documentclass[letterpaper,twocolumn,10pt]{article}
\usepackage{usenix}

\newcommand{\secref}[1]{\mbox{\S\hspace{0.1em}\ref{#1}}}

\newcommand{\figref}[1]{Fig.~\ref{#1}}

\newcommand{\minisection}[1]{\noindent\textbf{#1}}

\newcommand{\idot}{\raisebox{0.2ex}{\scalebox{0.85}{$\bullet$}}\thinspace}

\definecolor{textred}{RGB}{215,25,28}
\definecolor{textgreen}{RGB}{26,150,65}

\newcommand{\textbfit}[1]{\textbf{\textit{#1}}}

\usepackage{amsmath}

\usepackage{amssymb}

\usepackage{soul}
\usepackage{algorithm}
\usepackage{algorithmicx}
\usepackage{algpseudocode}

\usepackage{tikz}
\newcommand*\circled[1]{\tikz[baseline=(char.base)]{
            \node[shape=circle,draw,inner sep=1pt, line width=0.6pt] (char) {\textbf{#1}};}}
            
\usepackage[T1]{fontenc}
\usepackage{xspace}

\usepackage{CJKutf8}
\usepackage{booktabs}
\usepackage{listings}
\usepackage{xcolor}
\usepackage{enumitem}
\usepackage{soul}
\usepackage[normalem]{ulem}
\usepackage{mathtools}

\usepackage[most]{tcolorbox}

\definecolor{codegray}{gray}{0.95}
\definecolor{commentgreen}{rgb}{0,0.5,0}
\definecolor{keywordblue}{rgb}{0.0,0.0,0.7}

\begin{document}

\date{}

\newcommand{\sys}{Agent.xpu\xspace}

\title{\Large \bf \sys: Efficient Scheduling of Agentic LLM Workloads on Heterogeneous SoC}

\author{
\rm{Xinming Wei$^{\text{1}}$ \enskip
    Jiahao Zhang$^{\text{1}}$ \enskip
    Haoran Li$^{\text{1}}$ \enskip
    Jiayu Chen$^{\text{1}}$ \enskip
    Haoning Guan$^{\text{2}}$  \enskip
    Rui Qu$^{\text{1}}$ \enskip
    }
\\
\rm{
    Maoliang Li$^{\text{1}}$ \enskip
    Xiang Chen$^{\text{1}}$ \enskip
    Guojie Luo$^{\text{1,3}}$ \enskip
    }\\
\\
{$^{\text{1}}$School of Computer Science, Peking University\enskip $^{\text{2}}$The University of Hong Kong}\\ {$^{\text{3}}$National Key Laboratory for Multimedia Information Processing, Peking University\enskip}
}

\maketitle

\begin{abstract}

Personal LLM agents increasingly combine foreground reactive interactions with background proactive monitoring, forming long-lived, stateful LLM flows that interleave prefill and token-by-token decode. While modern heterogeneous SoCs integrate CPUs, iGPUs, and NPUs to support on-device intelligence, existing LLM engines assume static, single-shot inference and lack mechanisms for flow-level concurrency, prioritization, and efficient accelerator coordination. As a result, commodity SoCs remain poorly matched to the dynamic, mixed-criticality execution patterns of personal agents.

This paper presents \sys, the first LLM engine that orchestrates concurrent reactive and proactive LLM flows on commodity SoCs. Extensive profiling uncovers unique SoC characteristics of operator-accelerator affinity, asymmetric DDR contention, and stage-divergent batching behaviors distinct from cloud-serving assumptions. \sys introduces three key techniques: a heterogeneous execution graph (HEG) capturing NPU/iGPU affinity and elastic operator binding; flow-aware NPU-iGPU coordination with stage elasticity, decoupling prefill and decode to reduce bandwidth contention and enforce priorities; and fine-grained preemption with slack-aware piggybacking to guarantee reactive responsiveness without starving proactive work. Across realistic personal-agent workloads, \sys delivers 1.2-4.9$\times$ proactive throughput and reduces reactive latency by at least 91\%, compared with both industrial iGPU-only serving engine and NPU-iGPU static inference with optimal tensor-partitioning schemes. \sys also minimizes energy consumption and graphics interference via controlled iGPU usage.

\end{abstract}
\section{Introduction}

\begin{figure}[t]
    \centering
    \includegraphics[width=\columnwidth]{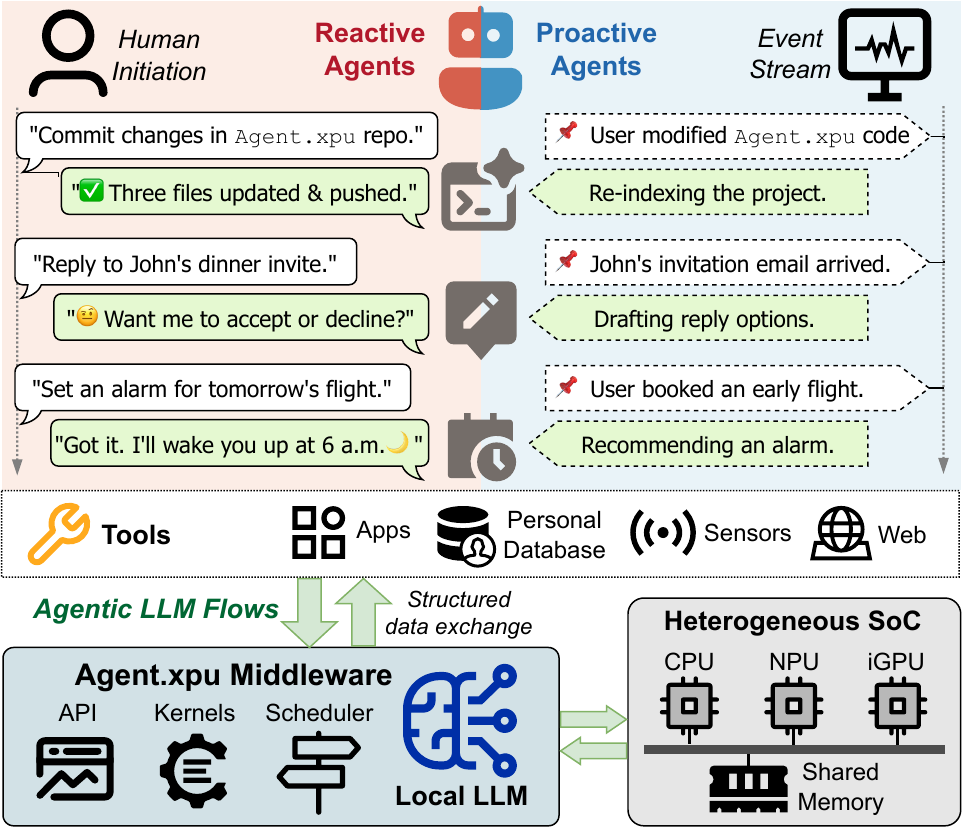}
    \caption{\textbf{Personal LLM Agent System.} \sys bridges the agent applications and heterogeneous SoC, orchestrating stateful on-device LLM flows from both foreground \textit{reactive} agents and background \textit{proactive} agents.}
    \label{fig:agent-workload}
\end{figure}

Large Language Models (LLMs) have revolutionized intelligent personal assistants~\cite{de2020intelligent}, powering autonomous agentic workflows that combine answering, planning, and tool interaction~\cite{li2024personal, wang2024survey, yang2024if}. Personal agents operate through two complementary modes~\cite{de2020intelligent, liao2023proactive, deng2024towards, ambient-agent, lu2024proactive}. As shown in \figref{fig:agent-workload}, \textit{reactive agents} respond to user-initiated queries in the foreground. Meanwhile, \textit{proactive agents} continuously monitor event signals and perform speculative analysis without explicit user triggers. The agentic execution paths intersect at \textit{LLM flows}, i.e., structured, stateful LLM invocations punctuated by stalls of human thinking or tool calling. Since LLM flows frequently manipulate personal data and involve interactive steps, on-device deployment is preferable for privacy, responsiveness, and cost efficiency~\cite{yuan2024mobile, yin2024llm, xu2024device}. Emerging 0.6B-8B lightweight LLMs~\cite{llama-3.2, phi-4-mini, yang2025qwen3, guo2025deepseek} finetuned for agentic behaviors increasingly act as on-device controllers for function calls~\cite{erdogan2024tinyagent}, with complex reasoning tasks selectively routed to cloud LLMs for effective edge-cloud collaboration~\cite{shao2025division}.

Modern heterogeneous system-on-chip (hetero-SoC) across laptops and mobile phones comprises CPU, integrated GPU (iGPU), and NPU, to support local generative AI. While recent studies of on-device LLM acceleration~\cite{mllm-npu, powerinfer-2, heterollm, heteroinfer} exploit hetero-SoC via quantization or NPU offloading, they basically target \textbfit{static inference}, assuming isolated, single-shot model executions with one-off prompts, non-overlapping scheduling, and uniform latency goals. In contrast, agentic LLM flows break these assumptions: they require \textbfit{flow-level concurrency, coordination, and prioritization}. Reactive flows contain real-time, bursty requests that demand timeliness; proactive flows comprise long-running, best-effort background tasks; and flow stages interleave unpredictably as agents reason or stall on external events. Current hetero-SoC hardware-software stacks offer no native support for such dynamic, flow-oriented execution, widening their mismatch with real-world personal agents.

We identify three fundamental gaps between efficient agentic LLM flows and current hetero-SoC ecosystems~(\secref{subsec:bg-on-device}).
\begin{itemize}[leftmargin=0.8em, labelsep=0.3em]
\item \textit{Flexibility-efficiency trade-off.} NPUs deliver high performance for static-shaped input and ahead-of-time-compiled computation graphs, yet they are ill-suited for variable sequence lengths, dynamic batch sizes, and irregular control flow inherent to agentic LLM workloads. Conversely, more flexible iGPUs suffer from lower energy efficiency and potential interference with graphics responsibilities.

\item \textit{Shared-memory contention.} Concurrent agentic flows often emit overlapping NPU and iGPU kernel executions, which intensify contention for the limited DDR bandwidth of modern shared-memory SoCs, potentially degrading the latency of each individual task.

\item \textit{Absence of flow-aware runtime abstractions.} Most on-device LLM engines lack mechanisms such as in-flight batching, prioritized scheduling, and coordinated NPU-iGPU control, which are essential for launching concurrent reactive and proactive flows while meeting their distinct latency or throughput requirements.
\end{itemize}

Despite the above constraints, our in-depth hetero-SoC profiling~(\secref{sec:soc-profile}) reveals concrete opportunities to accommodate agentic LLM flows efficiently on commodity SoCs. These insights are distinct from cloud-centric LLM serving assumptions, motivating the design of our system.
\begin{itemize}[leftmargin=0.8em, labelsep=0.3em]

\item \textit{Operator-accelerator affinity.} We observe distinct efficiency trade-offs between NPU and iGPU handling static versus dynamic LLM operators during prefill or decode, two LLM inference stages. This motivates both pre-run static placement and runtime elastic binding to map operators to their most efficient hardware backend.

\item \textit{Asymmetric contention pattern.} Memory-bound kernels are significantly more sensitive to, and more likely to saturate, shared DDR bandwidth than compute-bound kernels. This suggests that contention-aware kernel dispatching can maximize aggregate throughput by preventing concurrent memory-heavy accesses.

\item \textit{Stage-decoupled scheduling.} Co-locating all inference stages inevitably causes interference between reactive and proactive flows. However, distributing workloads based on distinct characteristics of prefill and decode across NPU and iGPU enables prioritized scheduling for reactive responsiveness while preserving proactive throughput.

\item \textit{Stage-divergent batching effects.} Prefill gains little from batching because it saturates compute regardless of batch size; decode batching improves multi-flow throughput, yet the individual latency of each flow is sensitive to total batch size and input length of all flows. This necessitates adaptive batching strategies for mixed reactive-proactive flows. 
\end{itemize}

\minisection{\sys Framework.} We present \sys, the first LLM engine that orchestrates agentic flows on commodity hetero-SoCs~(\figref{fig:agent-workload}). \sys not only fills the vacuum of fully on-device concurrent serving, but aligns heterogeneous accelerators with agentic objectives by jointly optimizing reactive timeliness, proactive throughput, and energy efficiency. Its design combines profiling-driven graph construction with online, flow-aware NPU-iGPU co-scheduling to balance these competing goals~(\secref{subsec:method-overview}).

\minisection{Key Techniques.} First, \sys introduces a \uline{heterogeneous execution graph (HEG)} that abstracts NPU/iGPU operator behavior in LLM flows, encoding affinity-guided placement constraints, elastic chunked kernel binding, and predictive performance annotations for runtime scheduling~(\secref{subsec:method-heg}). Second, \sys deploys \uline{flow-aware NPU-iGPU coordination with stage elasticity}: prefill flows opportunistically interleave NPU and iGPU executions, while decode flows remain iGPU-resident and continuously batched. This decoupled scheduling mitigates bandwidth contention. Dedicated iGPU arbitration, prefill compute partitioning, and on-the-fly NPU kernel warm-up dynamically redistribute work under fluctuating flow concurrency~(\secref{subsec:method-pd}). Third, \sys provides \uline{fine-grained preemption} for reactive queries across both stages, leveraging unified memory space for copy-free context switching~(\secref{subsec:method-preempt}). \sys also utilizes \uline{slack-aware piggybacking} with adaptive decode batching to prevent proactive starvation while preserving reactive responsiveness~(\secref{subsec:method-piggyback}).

\minisection{Implementation and Evaluations.} We implement both the agent interface and the LLM backend from scratch on Intel Core Ultra SoCs~\cite{intel-core-ultra}, representative of mainstream heteo-SoC architecture~(\secref{sec:impl}). \sys builds atop Intel OpenVINO~\cite{openvino} for efficient NPU/iGPU kernel implementations and hardware-native asynchronous APIs. \sys is orthogonal to low-bit quantization or sparse attention and therefore preserves model accuracy.

\sys substantially outperforms existing on-device LLM engines~(\secref{sec:eval}) across realistic personal-agent workloads (e.g., event handling, function calling, retrieval-augmented generation), using Llama3-3B/8B~\cite{llama-3.2} as representative on-device LLM backbones. Baselines comprise (a) OpenVINO iGPU \textit{serving}~\cite{openvino}, (b) Llama.cpp CPU \textit{serving}~\cite{llama.cpp}, and (c) customized \textit{serial} NPU-iGPU inference~\cite{heterollm, heteroinfer} with tuned tensor partitions for the target platform~(\secref{subsec:eval-setup}). Under proactive-only workloads, \sys yields 1.2-2.4$\times$ throughput over OpenVINO~(iGPU) and 1.4-4.9$\times$ over serial NPU-iGPU inference. For mixed reactive-proactive flows with diverse combinations of request heaviness, \sys reduces reactive query latency by 91-97\% while improving proactive throughput by 0.3-58\% relative to OpenVINO~(iGPU), with even larger gains over other baselines. The speedups are generally more pronounced on 8B than 3B model, indicating robust scalability~(\secref{subsec:eval-e2e}). \sys further reduces energy consumption by 26.8\% and iGPU utilization by 32.5\% compared with iGPU serving and NPU-iGPU serial baselines, respectively~(\secref{subsec:eval-overhead}).

\section{Background}

This section presents an overview of personal LLM agents~(\secref{subsec:bg-agent}), the basics of LLM inference~(\secref{subsec:bg-infer}), together with the landscape of hardware/software for on-device LLM, and their gaps towards smooth agentic flows~(\secref{subsec:bg-on-device}).

\subsection{Personal LLM Agents}\label{subsec:bg-agent}

In this work, we focus on \textit{personal} LLM agents deeply coupled with personal data, personal devices, and personal applications~\cite{li2024personal}. Analogous to the kernel in a traditional OS, the foundational LLM serves as the core execution engine of a personal LLM agent system, handling diverse queries issued by both reactive and proactive agents. Reactive agents are instantiated in direct response to explicit human requests. In contrast, proactive agents operate in a ``human-in-the-loop''~\cite{mosqueira2023human} manner, initiating actions autonomously based on contextual cues from the user’s environment, yet seeking human feedback or approval before execution. Mainstream LLM agent orchestration frameworks, such as LangChain~\cite{langchain}, LlamaIndex~\cite{llamaindex}, and AutoGen~\cite{autogen}, are equipped to customize both proactive and reactive agentic workflows.

\subsection{LLM Inference Primer}\label{subsec:bg-infer}

LLM inference is typically based on decoder-only Transformers, comprising a \textit{prefill} stage, which encodes the prompt and generates the first token, and an auto-regressive \textit{decode} stage to produce subsequent tokens one by one. Intermediate states (known as \textit{KV cache}~\cite{pope2023efficiently}) are updated after each step. The end-to-end latency of LLM inference consists of \textit{time to first token (TTFT)}, i.e., the prefill phase, and \textit{time per output token (TPOT)} multiplied by the number of decoded tokens after the first. In both prefill and decode, each Transformer layer constitutes three major blocks: \textit{1) QKV projection}, \textit{2) multi-head attention (MHA)}\footnote{We use MHA as a unified term to denote attention variants such as grouped-query attention (GQA)~\cite{ainslie2023gqa} and multi-latent attention (MLA)~\cite{liu2024deepseek-v2}.}, and \textit{3) feed-forward network (FFN)}. QKV projection and FFN operate independently on each \textit{token}, while attention attends over the entire \textit{sequence}. This distinction underlines differences in data dependencies and parallelism between the two types of operations.

Cloud-based LLM inference is dedicated to serving user queries at high throughput while meeting service level objectives (SLOs). Established techniques include kernel optimization~\cite{flashattention, smoothquant}, continuous batching~\cite{orca, vllm} with chunked prefill~\cite{sarathi}, prefill-decode disaggregation~\cite{distserve, splitwise}, KV cache reuse~\cite{sglang, preble, infercept} or defragmentation~\cite{vllm}, and improved tensor or pipeline parallelism~\cite{alpaserve, megatron}. For resource-constrained conditions where GPU memory is limited, existing solutions propose CPU memory/disk offloading by leveraging weight locality~\cite{powerinfer} or I/O-aware scheduling~\cite{flexgen}.

\subsection{On-Device LLM}\label{subsec:bg-on-device}

\begin{figure}
    \centering
    \includegraphics[width=\columnwidth]{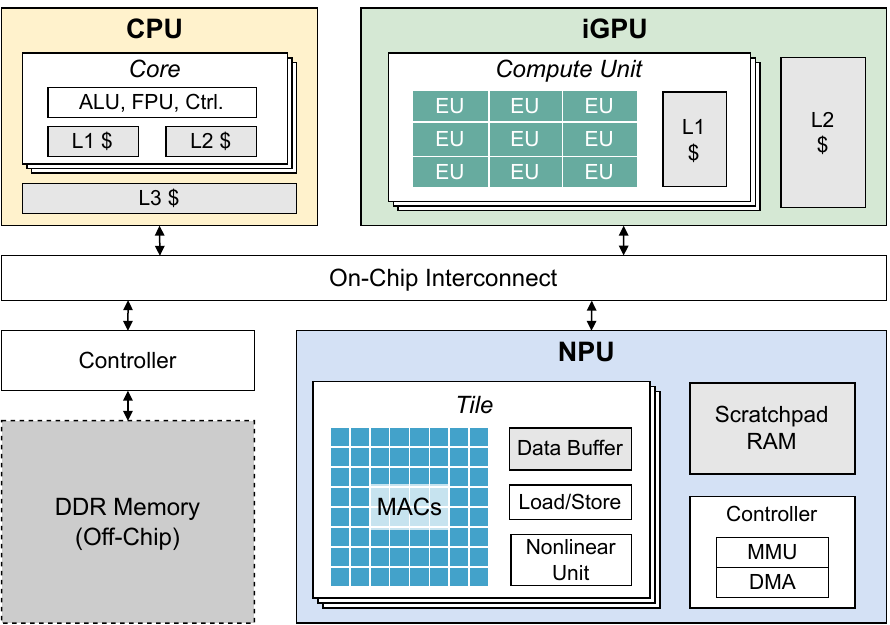}
    \caption{\textbf{Shared-Memory Hetero-SoC.} iGPU builds upon thread-level execution unit (EU), while NPU adopts multiply-accumulate (MAC) array for efficient tensor operations.}
    \label{fig:hetero-soc}
\end{figure}

\minisection{Heterogeneous SoC.} 
As illustrated in \figref{fig:hetero-soc}, hetero-SoCs exhibit a unified memory architecture distinct from heterogeneous systems with discrete accelerators: CPU, iGPU, and NPU share the same physical memory. This eliminates costly host-device data transfers but can introduce bandwidth contention caused by concurrent DDR access. Compared with cloud-hosted accelerators, commodity NPUs and iGPUs have smaller on-chip SRAM and depend heavily on shared, bandwidth-limited DRAM, requiring dedicated tuning of kernel chunking, batch dimension, and execution ordering.

From the compute angle, iGPU adopts a SIMT execution model akin to discrete GPUs, while the NPU is purpose-built for specific tensor operations, offering similar levels of parallelism but with higher energy efficiency. Nevertheless, LLMs process arbitrarily sized user inputs via dynamic-shape operators along the sequence or batch dimension, but NPUs are optimized for static-shaped operations. NPUs depend on costly compilation of fixed computational graphs to pre-allocate resources and optimize dataflows (e.g., tiling GEMMs onto fixed-size MAC arrays), making cold-start compilation at runtime infeasible. These trade-offs require NPU-iGPU collaboration to fully leverage the NPU's compute efficiency and minimize iGPU overhead.

\minisection{On-Device Inference Engines.} 
Driven by increasing demands for privacy, responsiveness, and energy efficiency, a range of industrial on-device LLM inference engines have emerged, including Llama.cpp~\cite{llama.cpp}, OpenVINO~\cite{openvino}, ONNX Runtime~\cite{onnxruntime}, IPEX-LLM~\cite{ipex-llm}, MNN~\cite{mnn}, QNN~\cite{qnn}, Core ML~\cite{coreml}, LiteRT~\cite{litert}, and MLLM~\cite{mllm-npu, yi2023mllm}. These frameworks facilitate out-of-the-box LLM deployment on vendor-specific CPUs, GPUs, or NPUs. Table~\ref{tab:framework-cmp} summarizes mainstream frameworks in comparison to \sys. \sys implements W8A16 quantization on both NPUs and GPUs to balance inference efficiency and accuracy precision, and supports stage-elastic NPU-iGPU hybrid inference. In contrast, Llama.cpp~\cite{llama.cpp} enables layer-wise CPU-iGPU pipelining without temporal overlapping. Llama.cpp and OpenVINO~\cite{openvino} support server-client-based LLM serving with continuous batching, which is limited on singular accelerator~(\secref{subsec:task-profile}). Only \sys supports preemptive scheduling during each inference stage, tailored for reactive flows with sub-100 ms wait time. Recent research has explored iGPU-NPU (or CPU-NPU) co-execution via activation outlier isolation~\cite{mllm-npu, powerinfer-2}, as well as tensor parallelism through activation or weight partitioning~\cite{heterollm, heteroinfer}. However, these techniques primarily target reducing the end-to-end latency of monolithic inferences, whereas future on-device agentic applications demand broader support for concurrency, state dependency, and real-time interactivity.

\begin{table}[t] \fontsize{7}{9}\selectfont

\caption{\textbf{Comparison of On-Device LLM Engines.}}

\begingroup
    \setlength{\tabcolsep}{0.36\tabcolsep}
\begin{tabular}{c|ccccc}
\hline
\textbf{Framework} &
  \textbf{iGPU} &
  \textbf{NPU} &
  \textbf{\begin{tabular}[c]{@{}c@{}}Hetero.\\ Execution\end{tabular}} &
  \textbf{\begin{tabular}[c]{@{}c@{}}Model\\ Serving\end{tabular}} &
  \textbf{\begin{tabular}[c]{@{}c@{}}Preempt.\\ Support\end{tabular}} \\ \hline\hline
Llama.cpp~\cite{llama.cpp}    & W4A16     & /      & CPU-iGPU & {\color[HTML]{CD9934} Cont. Batching} & / \\ \hline
ONNX Runtime~\cite{onnxruntime} & FP16/32 & INT8   & /                              & {\color[HTML]{656565} N/A (Offline)}  & / \\ \hline
IPEX-LLM~\cite{ipex-llm}     & W8A16     & INT4   & /                              & {\color[HTML]{656565} N/A (Offline)}  & / \\ \hline
MNN~\cite{mnn}          & W4A16     & /      & /                              & {\color[HTML]{656565} N/A (Offline)}  & / \\ \hline
MLLM~\cite{mllm-npu, yi2023mllm}         & /         & INT4   & /                              & {\color[HTML]{656565} N/A (Offline)}  & / \\ \hline
HeteroInfer~\cite{heteroinfer, heterollm}  & W4A16  & W4A16 & NPU-iGPU & {\color[HTML]{656565} N/A (Offline)} & / \\ \hline
Qualcomm AI~\cite{qualcomm-ai}  & W4A16     & INT4/8 & /                              & {\color[HTML]{656565} N/A (Offline)}  & / \\ \hline
OpenVINO~\cite{openvino}     & W8A16     & INT4   & /                              & {\color[HTML]{CD9934} Cont. Batching} & / \\ \hline
\textbf{\sys} &
  W8A16 &
  W8A16 &
  NPU-iGPU &
  {\color[HTML]{009901} \begin{tabular}[c]{@{}c@{}}Flow-Aware\\Decoupling\end{tabular}} &
  {\color[HTML]{009901} \begin{tabular}[c]{@{}c@{}}Stage-Aware\\Preemption\end{tabular}} \\ \hline
\end{tabular}
\label{tab:framework-cmp}
\endgroup

\begin{minipage}{\columnwidth}
\footnotesize
The listed iGPU/NPU quantization methods represent the most common configurations for each framework; CPU is omitted as typically supported by default. Both W8A16 and INT8 store weights in INT8, but use FP16 and INT activation/arithmetic, respectively; similarly for W4A16 and INT4.

\end{minipage}
\end{table}

\section{Hetero-SoC Analysis and Opportunities}\label{sec:soc-profile}

We conduct a comprehensive hetero-SoC analysis to guide the design of \sys. Operator-level analysis~(\secref{subsec:op-profile}) characterizes the compute and memory demands of representative LLM operators (ops), while task-level analysis~(\secref{subsec:task-profile}) examines runtime behavior in end-to-end agentic LLM flows. Profiling experiments are conducted on an Intel Core Ultra processor~\cite{intel-core-ultra} with DDR5 DRAM.

\subsection{Operator-Level Analysis}\label{subsec:op-profile}

\begin{figure}[t]
    \centering
    \includegraphics[width=\columnwidth]{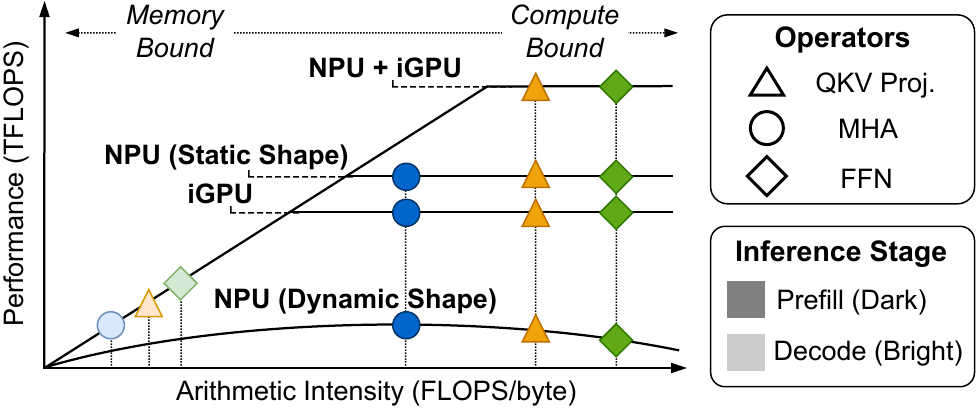}
    \caption{\textbf{Schematic Roofline Illustration of LLM Ops.}}
    \label{fig:profile-roofline}
\end{figure}

\begin{figure}[t]
    \centering
    \includegraphics[width=1.0\columnwidth]{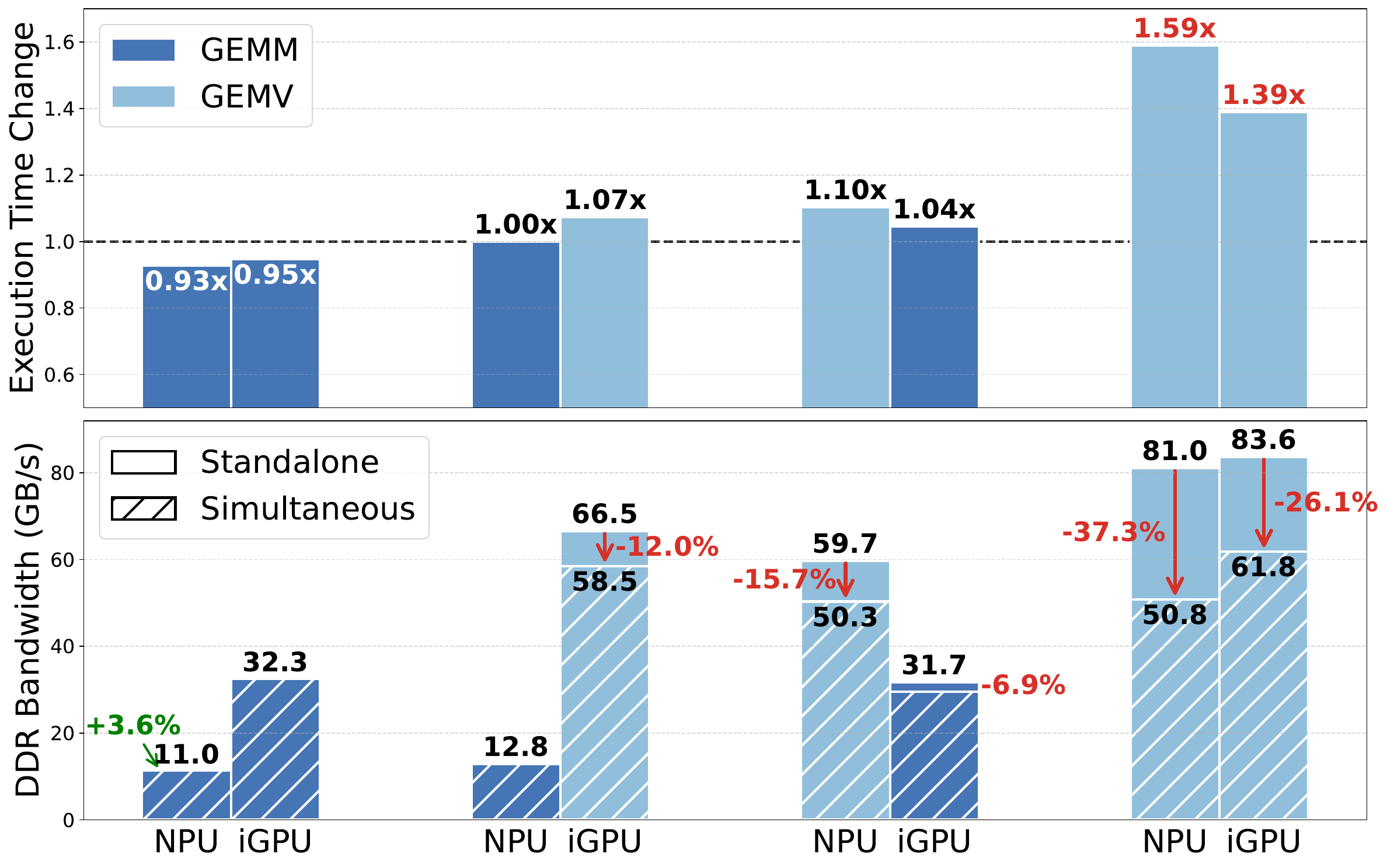}
    \caption{\textbf{Memory Contention Analysis.} Changes of execution time (upper) and DDR bandwidth (lower) from standalone NPU/iGPU kernel running to simultaneous co-execution. Memory-bound GEMV kernels are more sensitive to NPU/iGPU parallelism than compute-bound GEMM.}
    \label{fig:profile-contention}
\end{figure}

\minisection{Operator-Accelerator Affinity.} To illustrate the affinity between LLM ops and hetero-SoC accelerators, we construct a schematic roofline model (\figref{fig:profile-roofline}) derived from our profiling results of Llama 3B/8B models\footnote{Simplifications include: arithmetic intensity of each operator varies with prompt length (here is a moderate 512 tokens); roofline curves of different ops may actually diverge and are merged here; relative NPU/iGPU peak performance are vendor-dependent; memory bandwidth slope for each curve differs by factors such as L2 or scratchpad memory size; and NPU+iGPU parallelism can slightly boost bandwidth when memory-bound~(\figref{fig:profile-contention}, \cite{heteroinfer}).}. 
Static-shaped NPU kernels can be ahead-of-time (AOT) compiled, whereas dynamic-shaped kernels require expensive just-in-time (JIT) compilation. In contrast, iGPU natively supports AOT kernels with variable-shaped inputs. For dynamic prefill kernels on NPUs, we amortize JIT compilation cost by LLM layers across which they can be reused; nevertheless, this overhead remains prohibitive for on-demand runtime compilation. By comparison, static-shaped NPU kernels achieve higher energy efficiency (TFLOPS/W) while delivering throughput comparable to iGPUs. 
\textbfit{Opportunity:} Pre-compile chunked NPU kernels for token-wise prefill ops, while offloading dynamic prefill and all decode ops to iGPU, which better handles growing dimensions and avoids NPU underutilization~(\secref{subsec:method-heg}). At runtime, warm up variable-shaped kernels on NPU for queued requests on-the-fly to hide compilation latency and opportunistically shift prefill from iGPU~(\secref{subsec:method-pd}).

\minisection{Memory Access Pattern and Contention.}
Discrete GPUs with dedicated VRAM exhibit \textit{bulk transfer and decoupled compute}, moving data between host DRAM and VRAM before or after execution. In contrast, we observe that NPUs and iGPUs with limited SRAM and DMA capacity adopt \textit{streaming access and coupled compute}, consuming data progressively during execution, which underscores NPU-iGPU contention management under heavy DDR traffic. As shown in \figref{fig:profile-contention}, we measure latency and bandwidth changes between separate and simultaneous GEMM/GEMV executions with Intel VTune~\cite{vtune}. GEMM and GEMV dominate prefill and decode stages, respectively. Co-executing memory-bound GEMV degrades both latency and bandwidth, whereas compute-bound GEMM is largely unaffected. 
\textbfit{Opportunity:} Contention can be largely mitigated through stage decoupling of prefill and decode to NPU and iGPU~(\secref{subsec:task-profile}), complemented by adaptive kernel dispatching with deferral~(\secref{subsec:method-pd}).

\subsection{Task-Level Analysis}\label{subsec:task-profile}

\begin{figure}[t]
    \centering
    \includegraphics[width=\columnwidth]{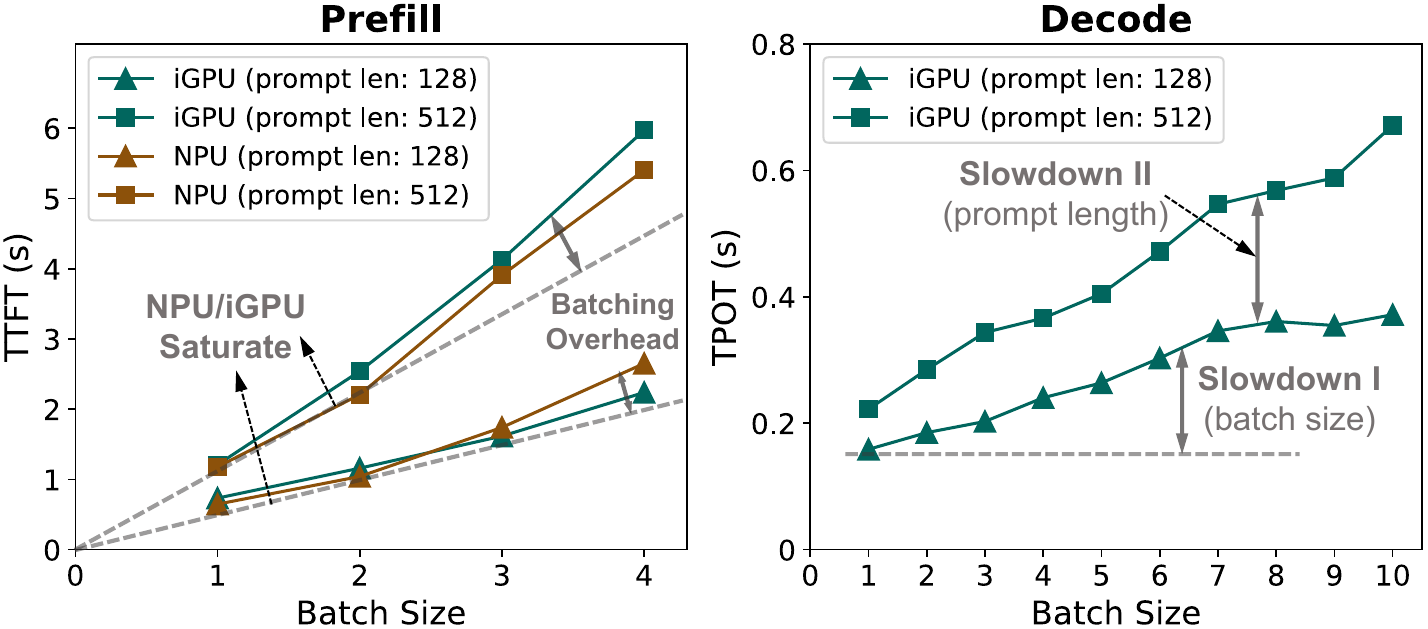}
    \caption{\textbf{Individual Task Latency in Batching.} Distinctive batching effects of prefill and decode on Llama-3B model.}
    \label{fig:profile-batch}
\end{figure}

\begin{figure}[t]
    \centering
    \includegraphics[width=\columnwidth]{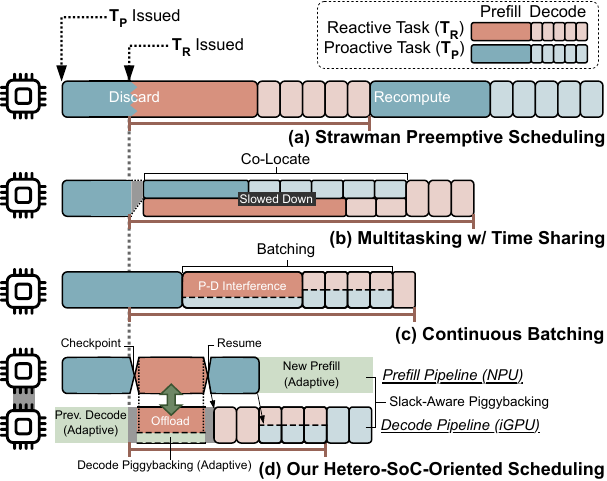}
    \caption{\textbf{Proactive-Reactive Co-Scheduling.} (a)(b)(c) target single accelerator, while (d) uses NPU and iGPU primarily for prefill and decode, respectively.}
    \label{fig:profile-co-schedule}
\end{figure}

\minisection{Batching Effects on Hetero-SoC.}
Batching intuitively improves throughput, while the individual latency is more sensitive to batch size on resource-constrained SoCs. As shown in \figref{fig:profile-batch}, we analyze prefill and decode batching across varying batch sizes, prompt lengths, and accelerators\footnote{NPU decode is omitted since per-iteration kernel compilation is infeasible under growing sequence length and varying batch size.}. Prefill latency scales nearly linearly with batch size, saturating the NPU or iGPU without performance gain. Decode latency rises gradually with larger batches or longer prompts, since each task's MHA operates on its own KV cache with unchanged arithmetic intensity, limiting batching efficiency. This effect is amplified by constrained memory bandwidth and quadratic MHA complexity in sequence length. Furthermore, the latency gap between prefill and decode reflects decode degradation when batched with prefill. \textbfit{Opportunity:} To trade-off latency and throughput in batching, we can disaggregate prefill and decode across NPU and iGPU~(\secref{subsec:method-pd}), and employ adaptive decode batching tailored to flow priorities~(\secref{subsec:method-piggyback}).

\minisection{Proactive-Reactive Interference.} 
Agentic LLM serves interleave proactive flows with real-time reactive flows, creating conflicting throughput-latency demands. \figref{fig:profile-co-schedule} compares four co-scheduling schemes. Single-accelerator approaches, including (a) instant preemption without context saving, (b) time-sharing via multi-stream or virtualization, and (c) continuous batching, all suffer inefficiencies: (a) recomputation and idle time, (b) slowdowns and duplicated intermediate buffers, and (c) prefill-decode interference, which lengthens decode and is exacerbated on resource-constrained SoCs with longer prefill times. By co-locating heterogeneous stages on the same accelerator with coupled resource allocation, these methods cannot simultaneously satisfy the objectives of agentic workloads.
Our hetero-SoC scheme (d) principally partitions prefill to the NPU and decode to the iGPU, with the iGPU also handling reactive prefill and dynamic prefill kernels. In this way, we achieve lowest reactive flow makespan and highest overall throughput. 
\textbfit{Opportunity:} Minimize reactive-proactive interference via efficient preemption for reactive tasks, plus recomputation-free resumption~(\ref{subsec:method-preempt}) and slack-aware piggybacking~(\secref{subsec:method-piggyback}) guided by batching and contention profiles for proactive tasks.

\section{\sys Design}\label{sec:method}

\begin{table}[t] \fontsize{7}{9}\selectfont
\centering
\caption{\textbf{Key Notations}}
\begin{tabular}{l|l}
\toprule
$\mathcal{Q}_{react}, Q_{proact}$ & Global request queues containing reactive/proactive tasks \\
$S_{len}(r)$ & Input sequence length of request $r$ \\
$S_{chk}$ & Fixed sequence length of a given AOT chunk kernel \\
$N_{chk}(r)$ & Total number of chunks for a request ($N_{chk}(r) = \lfloor S_{len}(r) / S_{chk} \rfloor$) \\
$n_{npu}$ & Number of chunks dispatched to NPU \\
$S_{rem}(r)$ & Remainder dynamic length ($S_{rem}(r) = S_{len}(r) \pmod {S_{chk}}$) \\
$buf_{p}$ & Prefill buffer with shape $(max\_context\_len, d_{model})$ \\
$buf_{d}$ & Decode buffer with shape $(max\_decode\_batch\_size, d_{model})$ \\
$\mathcal{B}_{dec}$ & Decode batch containing reactive or proactive requests \\
$\mathcal{D}_{status}$ & Status of iGPU Decode pipeline (IDLE or BUSY) \\
$\mathcal{T}_{op}(sz, dev)$ & Latency of a specific operator $op$ of input size $sz$ on device $dev$ \\

\bottomrule
\end{tabular}
\label{tab:notations}
\end{table}

We first summarize the objectives, workload assumptions, and essential components of our system~(\secref{subsec:method-overview}). Then we introduce the offline HEG~(\secref{subsec:method-heg}) and online NPU-iGPU coordination guidelines~(\secref{subsec:method-pd}). Finally, we elaborate on the reactive preemption~(\secref{subsec:method-preempt}) and proactive piggybacking~(\secref{subsec:method-piggyback}) mechanisms. Table~\ref{tab:notations} shows key notations used in this section.

\subsection{Contextualization and System Overview}\label{subsec:method-overview}

\begin{figure}[t]
    \centering
    \includegraphics[width=\columnwidth]{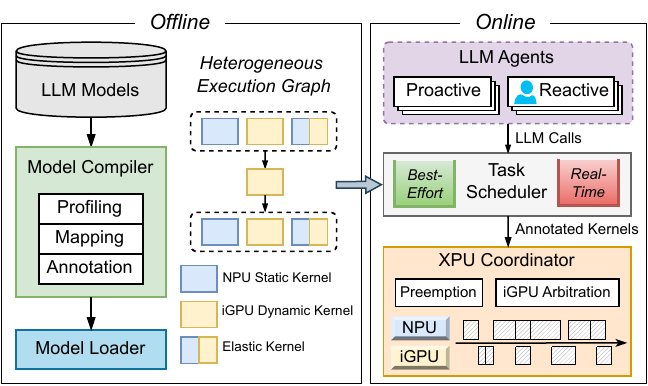}
    \caption{\textbf{\sys System Design.}}
    \label{fig:method-sys}
\end{figure}

\minisection{Role of \sys.} 
\sys serves concurrent LLM flows from personal agents, rather than handling standalone LLM inferences. It supposes one local LLM as the core of the agentic system. Operating in a non-clairvoyant manner, \sys does not rely on knowledge of the agentic workflow or task arrival times. It is informed only of task priorities (proactive or reactive) at the time of issuance. \sys is designed with the following primary \textbfit{objectives}: 1) prioritizing the reduction of end-to-end latency for LLM requests originating from reactive agents to enhance user experience, 2) increasing the overall throughput for background LLM flows from proactive agents, and 3) optimizing compute and memory resource utilization in hetero-SoC to achieve improved performance and energy efficiency.

\minisection{Workload Characteristics.} 
On resource-limited hetero-SoCs, we anticipate a typical LLM request rate from personal agents of 1 to 30 requests per minute. Calls from proactive and reactive agents are assumed to be independently distributed. LLM inferences are separate from other agentic sub-tasks, such as human interaction or tool engagement, which are assumed to be primarily CPU- or I/O-bound. Within \sys, iGPU usage is intentionally limited to ensure graphics availability and energy efficiency.

\minisection{System Design.} 
As depicted in \figref{fig:method-sys}, the system comprises offline and online components. 1) \textbfit{Offline}: it maintains model weights and a heterogeneous execution graph (HEG; \secref{subsec:method-heg}) with pre-compiled NPU kernels and dynamic iGPU kernels. Each HEG node carries profiling-guided annotations (latency and bandwidth hints, as a function of batch/sequence size). An elastic-kernel abstraction enables late binding of each operator to NPU or iGPU at dispatch time. 2) \textbfit{Online}: \sys dynamically schedules agentic LLM flows across the SoC on the basis of the following modules or data layouts:

\noindent \idot \textit{Request Manager.}
This module interfaces with the agent frontend to asynchronously admit LLM calls. It maintains a global request table, where each entry records the request UUID, lifecycle, KV cache allocation, prompt tokens, and inference progress (phase, layer, current kernel, generated tokens). This fine-grained tracking enables task checkpointing and resuming without recomputation. The manager handles lightweight admission and moves requests across task queues.

\noindent\idot \textit{Dual Task Queues.}
For both prefill and decode, \sys maintains a real-time queue (reactive) and a best-effort queue (proactive). The queues feed the corresponding event loops, enabling efficient scheduling and immediate preemption.

\noindent\idot \textit{Prefill/Decode Event Loops.}
Two dedicated loops busy-poll their queues for low-latency dispatch. They decompose tasks into NPU, iGPU, or elastic kernels and submit them to the XPU coordinator. Compute-bound prefill is run singly for each request, while decode loop adopts in-flight request batching~(\secref{subsec:method-pd}). The prefill loop supports sub-100 ms preemption for reactive requests at kernel boundaries~(\secref{subsec:method-preempt}), while the decode loop dynamically adjusts batch size based on request priority and latency budgets~(\secref{subsec:method-piggyback}).

\noindent\idot \textit{XPU Coordinator.}
This module orchestrates concurrent execution across NPU, iGPU, and related CPU activities (e.g., NPU kernel compilation). It operates on NPU/iGPU FIFO queues of submitted kernels, binding elastic kernels to accelerators and processing kernel-level preemption based on HEG annotations, task priority, and current load. The coordinator also arbitrates simultaneous iGPU requests under a prefill-prioritized policy~(\secref{subsec:method-pd}: \circled{3}).

\noindent\idot \textit{Recurrent Activation Double Buffer.}
\sys maintains a single-layer activation buffer, which can be reused recurrently through layers. Prefill and decode buffers are sized by max context length (set as 4096) and maximum batch size (set as 32), respectively, with the same hidden dimension. \sys adopts a reactive-proactive double buffer to enable copy-free context switching for preemption~(\secref{subsec:method-preempt}).

\noindent\idot \textit{Memory Manager.}
To fully utilize on-device memory, \sys employs a background garbage collector that reclaims KV caches and on-demand NPU kernels once completed. We assume moderate request density typical of personal agents without memory overflow. Should an out-of-memory condition arise, application-directed tiering is preferred over blind paging: selectively offloading cold KV cache or weight shards to flash storage. Such offloading policies are orthogonal to our core design and can be seamlessly integrated~\cite{flexgen}.

\subsection{Heterogeneous Execution Graph}\label{subsec:method-heg}

\begin{figure}[t]
    \centering
    \includegraphics[width=\columnwidth]{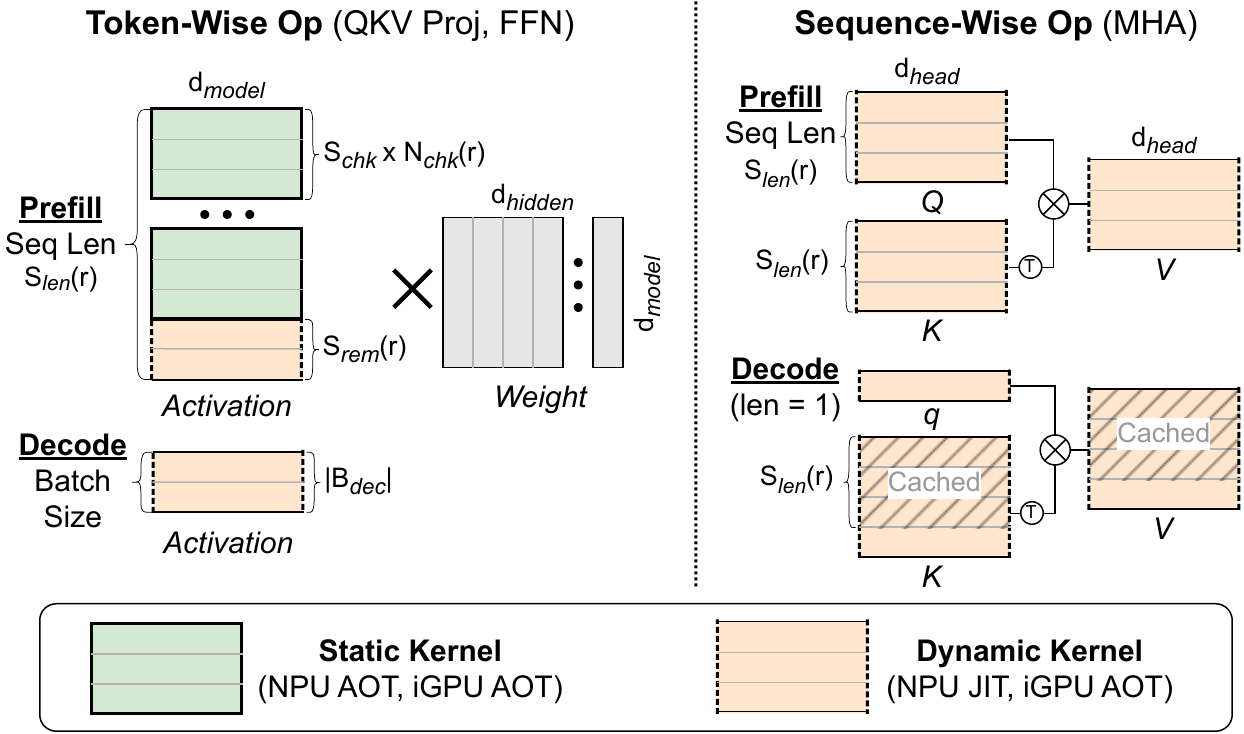}
    \caption{\textbf{HEG Op Decomposition and Elastic Binding.}}
    \label{fig:method-heg}
\end{figure}

\minisection{Graph Representation.}
\sys models LLM execution as a heterogeneous execution graph (HEG): a parametric, accelerator-aware multigraph where nodes denote operator variants and edges capture intra- or inter-accelerator data transfers. The HEG consists of a one-shot prefill DAG and a recurrent decode micro-DAG invoked per token. As illustrated in \figref{fig:method-heg}, token-wise QKV projection and FFN can be chunked during prefill or batched during decode with shared weights, while sequence-wise MHA operates per request over the individual KV cache.

\minisection{Operator Mapping.}
To map operators onto specific accelerators while enabling runtime elasticity, we balance computational efficiency, data locality, and resource utilization:
\begin{itemize}[leftmargin=0.8em, labelsep=0.3em]
 \item \textit{Computation Affinity.} Ops are placed based on roofline characteristics (\secref{subsec:op-profile}) and micro-architectural fit. For prefill, QKV projection and FFN use chunked NPU kernels or dynamic iGPU kernels; chunk sizes are tuned to saturate the NPU. Prefill MHA defaults to iGPU, though on-demand NPU warm-up can be employed. During decode, all kernels run on iGPU due to dynamic batch sizes, growing sequence lengths, and request heterogeneity. The CPU is reserved for control flow, toking sampling, and kernel compilation.
 \item \textit{Runtime Elasticity.} Prefill ops along the sequence dimension can be partitioned into chunked and dynamic parts. The coordinator adaptively chooses NPU or iGPU for each, depending on request priority, accelerator load, and bandwidth pressure. On-demand NPU kernels, if compiled in time, may also replace iGPU execution for dynamic ops.
 \item \textit{Memory Awareness.} To minimize costly DDR transfers, we fuse consecutive ops into three core kernels, exploiting NPU/iGPU SRAM with data locality. Unlike prior approaches~\cite{heterollm, heteroinfer, mllm-npu} that split linear and nonlinear ops across accelerators, our fused kernels exploit modern NPUs’ nonlinear units to avoid cross-accelerator transfer. For decode, we similarly fuse ops into three iGPU kernels, and further optimize by collapsing an entire decode layer into a single iGPU kernel for single-batch decode iteration.
\end{itemize}

\minisection{Predictive Kernel Annotation.}
Combining our roofline profiling~(\secref{subsec:op-profile}) and analytical model~(detailed in \secref{sec:perf-model}), we can predict kernel latency and bandwidth utilization as a function of sequence length or batch size, given accelerator choice. LLM ops are idempotent with fixed FLOPs, and we find that kernel execution times on NPU/iGPU are stable across invocations. This allows the scheduler to estimate TTFT for individual tasks and TPOT for batched decode, enabling adaptive kernel dispatching.

\subsection{Flow-Aware NPU-iGPU Coordination with Stage-Elasticity}\label{subsec:method-pd}

\sys principally decouples prefill and decode into specialized scheduling pipelines, facilitating the effective coordination of agentic flows. Prefill primarily leverages the NPU for compute-intensive token-wise ops, while decode resides on the iGPU to support dynamic batching and sequence growth. Such heterogeneous disaggregation is elastic to fit the real-time states of on-going reactive and proactive flows.

\minisection{\circled{1} Prefill Pipeline.}
For reactive requests, we minimize TTFT by partitioning token-wise ops across NPU and iGPU with elastic tensor parallelism (details in \circled{4}). For proactive requests, token-wise ops run mainly on NPU, while dynamic ops (e.g., MHA) and residual fragments alongside chunked kernels run on iGPU. Because a single request already saturates either accelerator, prefill avoids batching and processes requests serially. This pipeline also serves as the substrate for the preemption mechanism of reactive requests~(\secref{subsec:method-preempt}).

\minisection{\circled{2} Decode Pipeline.}
The decode pipeline adopts continuous batching decoupled from prefill, scheduling requests at iteration granularity. Reactive and proactive requests can be co-batched under adaptive strategies~(\secref{subsec:method-piggyback}). Executing all decode kernels on the iGPU naturally accommodates dynamic sequence lengths and fluctuating batch composition.

\minisection{\circled{3} Pipeline Coordination and iGPU Arbitration.}
On shared memory SoCs, prefill and decode share the in-place KV cache without costly cross-accelerator transfers. However, both stages may contend for the iGPU due to common dynamic operators, necessitating arbitration mechanisms. \sys adopts a \textit{prefill-first} arbitration policy: iGPU kernels from prefill always take precedence over decode, regardless of request type. This design is motivated by: 1) decode is memory-bound and generally longer than prefill, and 2) prefill requires only a small fraction of iGPU kernels, with the bulk of computation offloaded to the NPU. Prioritizing prefill ensures that short bursts of iGPU work complete promptly, avoiding long stalls that would otherwise block the entire prefill pipeline and inflate TTFT. Decode jobs can then efficiently utilize the wide gaps between prefill bursts.

\minisection{\circled{4} Elastic NPU-iGPU Tensor Parallelism.} 
To reduce TTFT for reactive requests while mitigating interference with ongoing decode, \sys elastically partitions reactive prefill kernels across NPU and iGPU at runtime. Decisions are made \textit{layer-wise} by the XPU coordinator in real-time. As detailed in Algorithm~\ref{alg:elastic_tp}, if decode pipeline is idle, the coordinator solves for $n_{npu}$ such that the NPU execution time for chunks aligns with the iGPU's execution time (assigned chunks + dynamic remainder), minimizing the makespan. Otherwise, the coordinator assigns all chunks to NPU to protect the latency-sensitive decode stream on iGPU. The remainder iGPU kernel is deferred to complete no earlier than their parallel NPU counterparts, increasing the chance that decode finishes mid-layer with interference alleviated.

\begin{algorithm}[t]
\footnotesize
\caption{Elastic Kernel Dispatch for Reactive Prefill}
\label{alg:elastic_tp}
\begin{algorithmic}
\Require Reactive request $r$ with input tokens $x$ (shape: $S_{len}(r) \times d_{model}$), layer $l$, prefill buffer $buf_{p}$ and decode status $\mathcal{D}_{status}$
\Ensure Completed execution of layer $l$ during prefill of $r$.
\If{$l$ is the first layer}
    \Comment{Fill prefill buffer with $r$'s input}
    \State memcpy($buf_{p}$, $x$, $S_{len}(r)\cdot d_{model}\cdot\text{sizeof}(x.dtype)$)
\EndIf
\For{each $op \in \text{GetOperators}(l)$}
    \If{$op$ is \textit{token-wise (QKV Proj, FFN)}}
        \If{$\mathcal{D}_{status}\text{ == }\text{IDLE}$}
            \Comment{Scenario 1: Maximize parallelism}
            
            \State $n_{npu} \gets \text{argmin}_{0 \le i \le N_{chk}(r)} \lvert i \cdot \mathcal{T}_{op}(S_{chk},\text{NPU})-$\\
            \hspace{8.0em}$(N_{chk}(r)-i) \cdot \mathcal{T}_{op}(S_{chk},\text{iGPU}) - \mathcal{T}_{op}(S_{rem}(r), \text{iGPU}) \rvert$
        \Else
            \Comment{Scenario 2: Conservative iGPU usage}
            \State $n_{npu} \gets N_{chk}(r)$
            \State $t_{defer} \gets n_{npu} \cdot \mathcal{T}_{op}(S_{chk},\text{NPU}) - \mathcal{T}_{op}(S_{rem}(r),\text{iGPU})$
        \EndIf
        \For{$i=0,\ldots,n_{npu}-1$}
            \Comment{Non-blocking}
            \State $\text{LaunchKernelAsync}(op, buf_{p}[i\cdot S_{chk}:(i+1)\cdot S_{chk}], NPU)$
        \EndFor
        \For{$i=n_{npu},\ldots,N_{chk}(r)-1$} \Comment{Non-blocking}
            \State $\text{LaunchKernelAsync}(op, buf_{p}[i\cdot S_{chk}:(i+1)\cdot S_{chk}], iGPU)$
        \EndFor
        \State $\text{Sleep}(t_{defer})$ \Comment{Defer iGPU kernel}
        \State $\text{LaunchKernel}(op, buf_{p}[N_{chk}(r)\cdot S_{chk}:S_{len}(r)], iGPU,$\\
        \hspace{6.0em} $\text{preempt}=true)$ \Comment Preempt (probable) decode kernels
        \State $\text{SyncExecution}(op, \text{NPU}, \text{iGPU})$
    \Else \Comment{Sequence-wise (MHA)}
        \State $\text{LaunchKernel}(op, buf_{p}[0:S_{len}(r)], iGPU, \text{preempt}=True)$
    \EndIf
\EndFor

\end{algorithmic}
\end{algorithm}

\minisection{\circled{5} On-the-Fly NPU Kernel Warm-Up.} 
At runtime, \sys opportunistically prepares dynamic NPU kernels (e.g., MHA) once a request is queued or preempted, thereby hiding compilation latency and reducing iGPU prefill load. Since prompt length is known at enqueue time, the CPU can start compiling static NPU kernels immediately; if compilation completes before prefill begins, the request switches to pure-NPU prefill, mitigating iGPU interference with decode. Compiled kernels are reclaimed once unused or expired. To eliminate potential contention introduced by NPU-iGPU co-execution, when NPU prefill kernels become memory-bound (e.g., MHA with short prompt length) and overlap with memory-intensive iGPU decode, the coordinator prioritizes reactive tasks: if both pipelines share the same priority (all-reactive or all-proactive), they proceed concurrently; otherwise, work with lower priority is deferred until the higher-priority side completes for reactive latency preservation.

\subsection{Fine-Grained Preemption}\label{subsec:method-preempt}

Prioritized scheduling of reactive flows is critical for user experience. However, preemption mechanism tailored to agentic LLM workloads is absent on commodity hetero-SoCs. We exploit the unified memory and our double activation buffer design to achieve lightweight context switching. Considering fine-grained, short-lived prefill kernels and moderate reactive request frequency, we adopt lazy preemption at kernel or layer boundaries without user-perceptible delay, which avoids recompute overhead in kill-based approaches~\cite{han2022microsecond}.

\minisection{\circled{1} Copy-Free Context Switching.} 
An inference context comprises the KV cache, progress metadata (stage, current layer, finished kernels, generated tokens), and intermediate activations. Shared memory address eliminates the need to swap the KV cache when a reactive request arrives. Our double activation buffer separates proactive and reactive activations of prefill, enabling low-latency preemption and instant restoration without data copying.

\minisection{\circled{2} Kernel-Level Prefill Preemption.} 
During prefill, we employ wait-then-execute preemption at kernel boundaries. Each kernel is registered with a lightweight completion callback. Upon finishing, the callback efficiently probes the reactive queue; if non-empty, it signals the prefill event loop to immediately dequeue and decompose the reactive job into kernels, thereby preempting the current proactive task. Reactive tasks do not preempt each other and run sequentially like proactive ones. Because token-wise ops are chunked, kernel execution time is tightly bounded. For an 8B model, chunked kernels (size 256) complete within 38 ms on NPU/iGPU, while MHA with sequence length 2048 completes within 97 ms. This ensures sub-100 ms preemption latency—nearly imperceptible to users with negligible TTFT impact.

\minisection{\circled{3} Iteration-Level Decode Preemption.} 
For decode, reactive requests wait until the current iteration completes, then immediately join the next batch. Proactive jobs may be evicted under adaptive batching policies~(\secref{subsec:method-piggyback}) to preserve reactive latency. Since iteration-level preemption is bounded by TPOT and the only per-request context is the localized KV cache and generated tokens, it avoids expensive swapping while still guaranteeing fast response.

\subsection{Slack-Aware Piggybacking}\label{subsec:method-piggyback}

To preserve reactive responsiveness while preventing starvation of proactive flows, \sys employs slack-aware piggybacking, where proactive work opportunistically fills pipeline slack in prefill and decode with negligible impact on reactive task latency.

\minisection{\circled{1} Slack Identification.} 
The scheduler detects two forms of slack during reactive-proactive coexistence:
1) \textit{Compute slack}, arising from idle accelerator resources. Proactive prefill can overlap with the decode pipeline whenever no reactive prefill is pending.
2) \textit{Bandwidth slack}, during memory-bound decode stage. Proactive requests can be co-batched with reactive ones under elaborate batching rules that bound the additional latency seen by reactive tasks, as \circled{2}.

\minisection{\circled{2} Adaptive Decode Batching.} 
\sys implements \textit{reactive-first} batching~(Algorithm~\ref{alg:adaptive_batch}) to govern the admission of proactive tasks into the next decode iteration. Since decode latency scales with both batch size and sequence length, blind mixing of requests can violate reactive latency targets. The scheduler updates the decode batch $\mathcal{B}_{dec}$ by unconditionally admitting all surviving and newly arrived reactive requests. Proactive requests are then opportunistically piggybacked into the remaining slots up to a threshold $B_{cap}$~(set as 3). Crucially, when the candidate set exceeds capacity, \sys evicts ``bottleneck'' proactive requests with the longest sequence lengths to preserve TPOT of the entire batch.

\begin{algorithm}[t]
\footnotesize
\caption{Reactive-First Adaptive Decode Batching}
\label{alg:adaptive_batch}
\begin{algorithmic}
\Require Current batch $\mathcal{B}_{dec}$, new arrivals $\mathcal{Q}_{react}, \mathcal{Q}_{proact}$, threshold $B_{cap}$
\Ensure Updated $\mathcal{B}_{dec}$ for the next decode iteration

\State $\mathcal{R}_{react} \gets \{ r \in \mathcal{B}_{dec} \mid r.\text{type}\text{ == }\text{REACTIVE} \} \cup \text{PopAll}(\mathcal{Q}_{react})$
\State $\mathcal{R}_{proact} \gets \{ r \in \mathcal{B}_{dec} \mid r.\text{type}\text{ == }\text{PROACTIVE} \} \cup \text{PopAll}(\mathcal{Q}_{proact})$

\State $\mathcal{B}_{dec} \gets \mathcal{R}_{react}$
\Comment{Unconditionally admit reactive requests}

\State $\text{SortDescending}(r \in \mathcal{R}_{proact}, \text{key}=S_{len}(r))$

\For{$r \in \mathcal{R}_{proact}$}
    \If{$|\mathcal{B}_{dec}| < B_{cap}$ or $\mathcal{R}_{react}\text{ == }\varnothing$}
        \State $\mathcal{B}_{dec} \gets \mathcal{B}_{dec} \cup \{r\}$
    \Comment{Piggyback proactive requests}
    \Else
        \State $\text{PushBackToQueue}(r, Q_{proact})$ \Comment{Evict to prevent TPOT inflation}
    \EndIf
\EndFor
\end{algorithmic}
\end{algorithm}

\minisection{\circled{3} Starvation Prevention.} 
While reactive requests receive priority, the scheduler implements aging mechanisms to prevent indefinite postponement of proactive requests. Proactive requests that exceed a threshold age are promoted to prevent starvation. They obtain two privileges without suspending reactive execution: 1) the scheduler reallocates iGPU to manage the overdue proactive prefill, while retaining reactive prefill (except MHA) solely on NPU, and 2) after prefill, the aged request can immediately join the decode batch, bypassing the batch size cap for a reactive-first batch.
\section{Implementation}\label{sec:impl}

\sys exposes a RESTful API frontend (2.3K lines of Python code) for prioritized query submission in a server-client manner. The LLM backend is written in 6K lines of C++ with custom abstractions for tensors, compute graphs, and model loading, eschewing third-party dependencies (e.g., PyTorch~\cite{pytorch}, GGML~\cite{ggml}). This lightweight design allows \sys to natively embed the scheduling mechanisms introduced in \secref{sec:method}, while keeping kernel interfaces modular to accommodate diverse SoC platforms.  

Our prototype targets Intel Core Ultra SoCs~\cite{intel-core-ultra}, implementing NPU and iGPU kernels using the low-level APIs of OpenVINO 2025.2~\cite{openvino}. Request- and kernel-level scheduling is based on two-tier asynchronous interfaces: 1) \textit{Inter-op parallelism} leverages thread-level prefill and decode event loops, coordinated via thread-safe task queues; 2) \textit{Intra-op parallelism} exploits elastic NPU-iGPU tensor partitioning, fulfilled by the XPU coordinator with hardware-specific coroutines (\texttt{start\_async/wait} in OpenVINO). The offline HEG and online scheduler jointly enable seamless adaptation to different LLM models, agent workflows, and SoC platforms.

\section{Evaluation}\label{sec:eval}

We evaluate \sys on diverse agentic flow against industrial baselines~(\secref{subsec:eval-setup}). Overall, \sys improves proactive throughput by up to 2.4$\times$ over iGPU serving baseline, while reducing reactive latency by up to 97\% under load~(\secref{subsec:eval-e2e}). Detailed breakdowns further attribute these gains to efficient heterogeneous co-scheduling~(\secref{subsec:eval-latency}). We also examine the overhead management~(\secref{subsec:eval-overhead}).

\subsection{Experimental Setup}\label{subsec:eval-setup}

\minisection{Hetero-SoC Testbed and Models.}
We deploy \sys on an ASUS NUC 14 Pro+ mini-PC equipped with an Intel Core Ultra 5 125H processor~\cite{intel-core-ultra} and 64GB DDR5 DRAM, running Ubuntu 24.04. The processor integrates an Intel Arc iGPU and Intel AI Boost NPU, with NPU driver v1.19.0, Intel iGPU Compute Runtime 24.39.31294.12, and performance power mode. We evaluate Llama-3.1-8B-Instruct and Llama-3.2-3B-Instruct~\cite{llama-3.2} as central LLM, covering representative model architecture (GQA, dense FFN) and sizes for on-device deployment. We adopt W8A16 channel-wise quantization, incurring negligible accuracy loss~\cite{Huang_2024}.

\begin{figure}[t]
    \centering
    \includegraphics[width=\columnwidth]{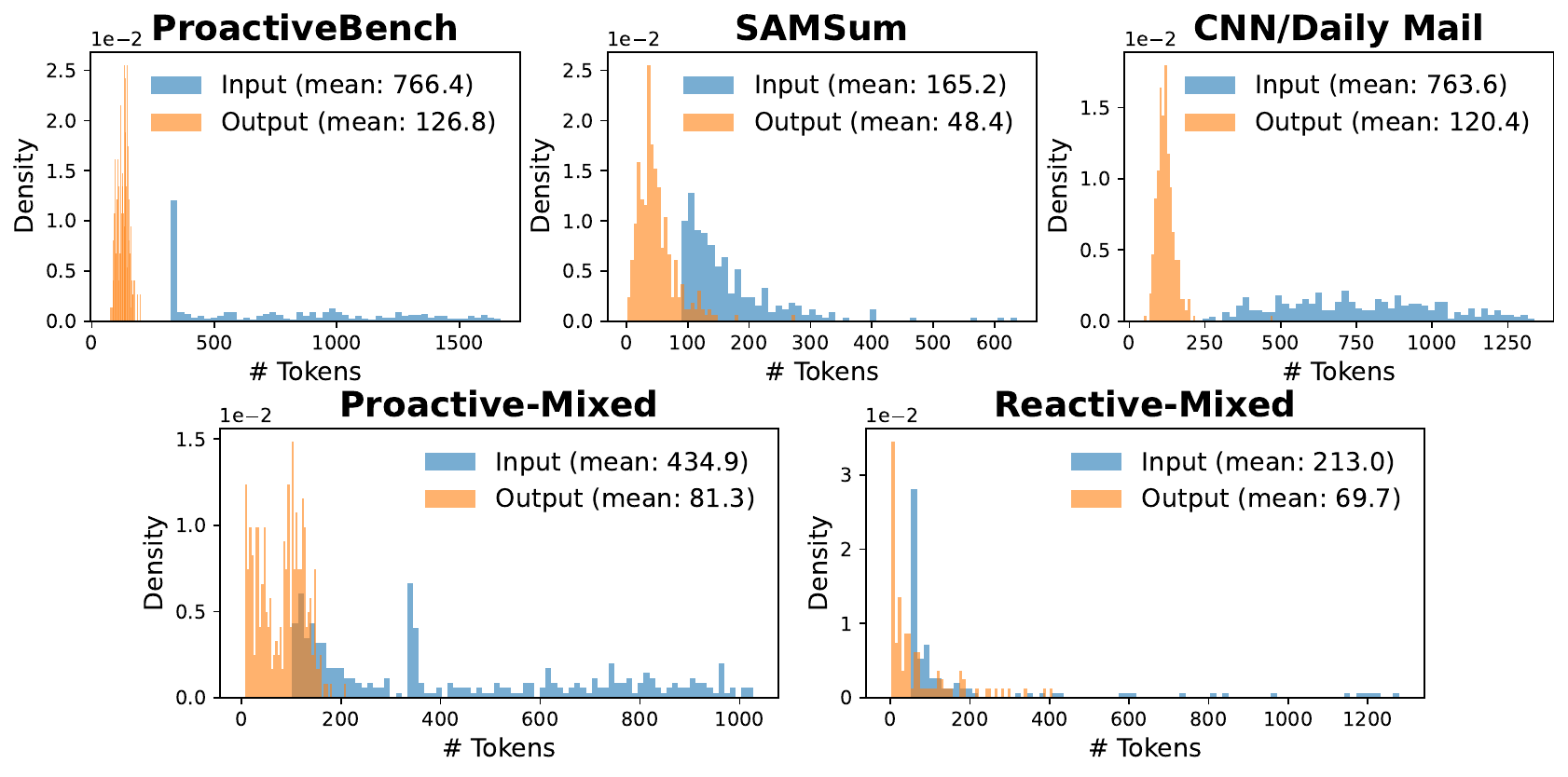}
    \vspace{-15pt}
    \caption{\textbf{Composition of Agentic Workloads.} Input/output length distributions of three proactive datasets and two proactive- or reactive-mixed datasets.}
    \label{fig:dataset}
\end{figure}

\minisection{Agentic Workloads.} To approximate realistic personal agent behavior, we construct both proactive and reactive workloads:  
\textit{Proactive:} 1) ProactiveBench~\cite{lu2024proactive} with real user events such as keyboard, clipboard, and browser activity; 2) SAMSum~\cite{Gliwa_2019}, modeling group chat reply drafting; and 3) CNN/DailyMail~\cite{hermann2015teach}, for news summarization. \textit{Reactive:} 1) LMSys-chat-1M~\cite{zheng2023lmsyschat1m}, covering diverse one-on-one conversations; 2) MTRAG~\cite{katsis2025mtrag}, a multi-turn retrieval-augmented generation benchmark; and 3) Berkeley Function Call Leaderboard~\cite{patil2023gorilla}, which produces structured API calls.

We evaluate two regimes:  
1)~\textit{proactive-only}: each proactive benchmark is run individually;  
2)~\textit{mixed flows}: proactive and reactive requests coexist. 
To synthesize mixed workloads, we uniformly sample from all proactive or reactive benchmarks to form proactive- or reactive-mixed datasets. 
As shown in \figref{fig:dataset}, the datasets exhibit remarkably different input-length distributions, reflecting the wide variation in prefill intensity across agentic tasks. Output lengths follow a long-tailed pattern, with outliers reaching up to 1.6k tokens—consistent with on-device agents producing summarized responses or compact function-call instructions, while more verbose reasoning tasks remain predominantly cloud-served.

Since the original datasets lack timestamps, we synthesize the arrival times using a Poisson process with request rates varying between 1-30 requests per minute (req/min), matching observed densities for personal agents. Proactive and reactive arrivals are generated independently. For each dataset, we evaluate the baselines and \sys with 15-minute traces, corresponding to increasing request rates.

\minisection{Baselines.}
We compare against industrial on-device LLM engines with \textit{concurrent-serving} support and optimizations for the underlying Intel SoC, as well as a customized NPU-iGPU \textit{static-inference} baseline. All baselines use the same W8A16 precision as \sys.
\begin{itemize}[leftmargin=0.8em, labelsep=0.3em]
    \item Llama.cpp~(CPU)~\cite{llama.cpp}, a widely used inference engine optimized for multi-core CPUs. We evaluate its serving mode with continuous batching.  
    \item OpenVINO~(iGPU)~\cite{openvino}, Intel’s deployment stack for Core Ultra SoCs. Since \sys also builds on OpenVINO’s low-level APIs, single-batch performance is nearly identical. For serving, OpenVINO supports continuous batching only on iGPUs, which we use as the iGPU-serving baseline.
    \item Serial~(NPU-iGPU). To generalize static tensor-parallel acceleration across NPU and iGPU~\cite{heteroinfer, heterollm}, we craft a serial pipeline that partitions prefill prompts into optimal NPU-iGPU ratios tuned for various lengths, and executes decode fully on the iGPU as NPU-iGPU parallelism barely yields decode benefit on our evaluated SoC.
\end{itemize}

\minisection{Metrics.}
We measure both performance and efficiency metrics under varying proactive and reactive request densities: 1)~\textit{Normalized Latency}, calculated as the average request end-to-end latency divided by combined input and output lengths, gauging throughput under high request rates. We also measure P90 latency of reactive requests to highlight user-facing responsiveness; 2)~\textit{iGPU Utilization}, the weighted sum of stable utilization percentage measured under distinctive loads or stages, averaged by the corresponding active execution periods; 3)~\textit{Energy per Token}, measured as the energy consumed (J/token), normalized by the processed token count.

\subsection{End-to-End Performance}\label{subsec:eval-e2e}

\begin{figure*}[ht]
    \centering
    \includegraphics[width=\textwidth]{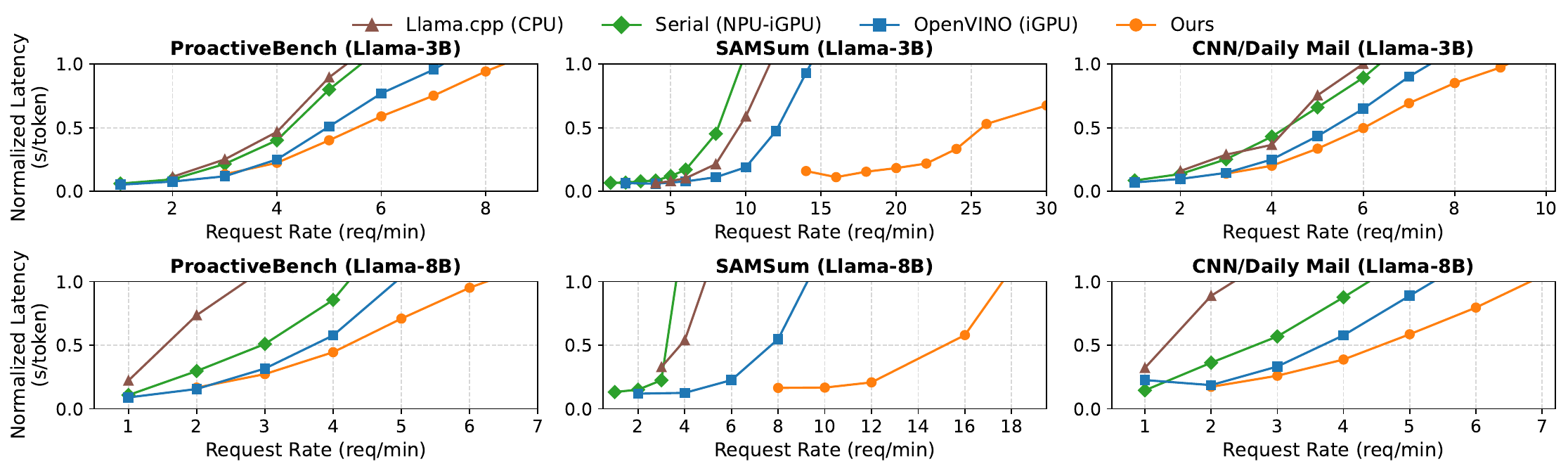}
    \caption{\textbf{Proactive-Only Workloads.} End-to-end results with Llama-3B/8B models on three proactive datasets.}
    \label{fig:eval-proact}
\end{figure*}

\begin{figure*}[ht]
    \centering
    \includegraphics[width=\textwidth]{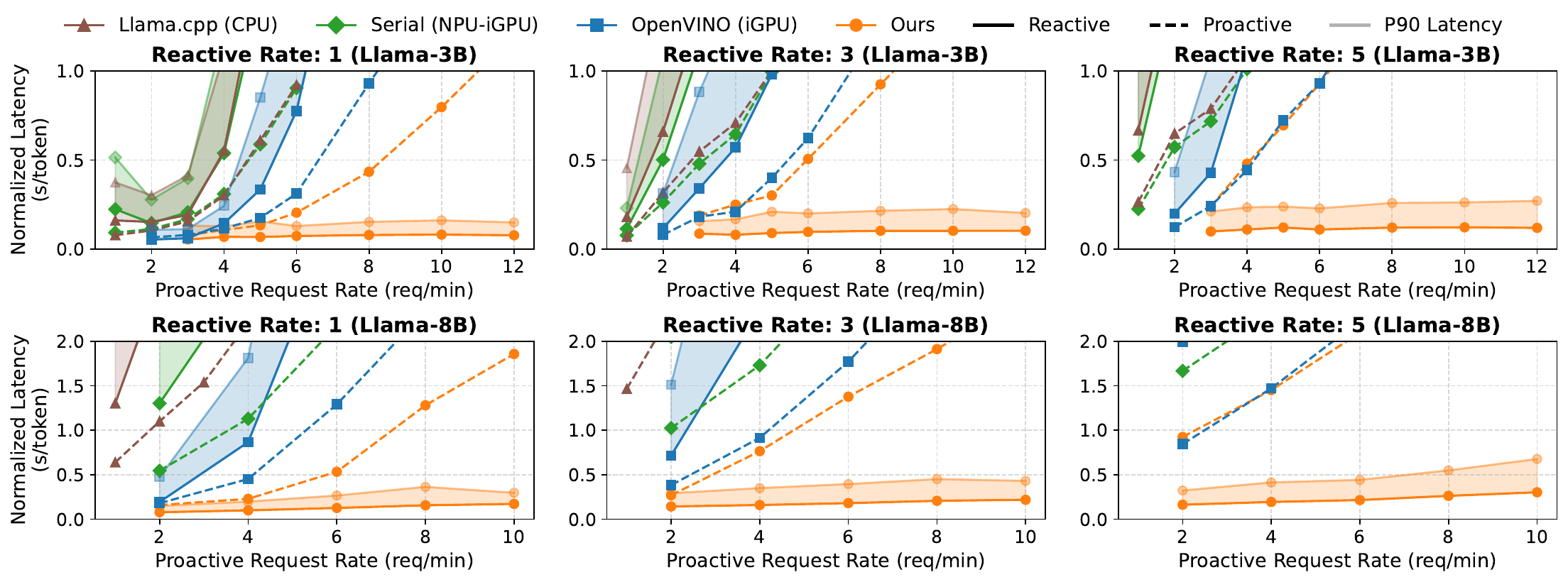}
    \caption{\textbf{Reactive-Proactive Mixed Workloads.} End-to-end results with Llama-3B/8B models on mixed datasets with varying reactive/proactive rate combinations. P90 latencies of reactive requests are displayed alongside mean latencies.}
    \label{fig:eval-mixed}
\end{figure*}

\minisection{Proactive-Only Workloads.}
We first examine proactive-only workloads, where throughput is the primary metric. In this regime, \sys assigns uniform priority without preemption or adaptive batching. Although designed for mixed workloads, \sys already outperforms other single- or dual-accelerator baselines in throughput~(\figref{fig:eval-proact}). OpenVINO (iGPU) ranks second, benefitting from throughput-oriented continuous batching. Llama.cpp~(CPU) excels in decode but is bottlenecked by slow prefill, while serial NPU-iGPU execution is limited by non-overlapping execution. \sys shares low-level kernels with OpenVINO with similar single-batch inference speed, but it achieves lower iGPU utilization~(\secref{subsec:eval-overhead}) by leveraging NPU-iGPU co-execution.

The performance advantage of \sys is further amplified under higher request rates or larger models. On SAMSum, where prompts and outputs are relatively short, all systems sustain higher rates before saturation. With Llama-3B, \sys achieves 2.0-2.4$\times$ throughput over OpenVINO (iGPU), and delivers $\sim$20\% more requests at 1.0 s/token latency on ProactiveBench and CNN/DailyMail. Gains are even larger with Llama-8B, as a heavier load amplifies the benefit of heterogeneous scheduling. Compared with Llama.cpp (CPU) and Serial~(NPU-iGPU), \sys achieves up to 3.9$\times$ and 4.9$\times$ throughput, respectively.


\minisection{Reactive-Proactive Mixed Workloads.}
We next evaluate mixed workloads, where proactive throughput and reactive latency must be balanced. Baselines lack request prioritization, so they treat proactive and reactive inputs equally. In contrast, \sys enforces reactive-first scheduling with adaptive batching. \figref{fig:eval-mixed} reports mean and P90 latencies to capture both responsiveness and tail behavior.

\sys consistently achieves much lower reactive latencies while sustaining higher proactive throughput. For Llama-3B at 6 proactive req/min, \sys reduces mean reactive latency by 91.61\%, 93.84\%, and 96.01\% at 1, 3, and 5 reactive req/min, respectively, compared to OpenVINO (iGPU). With Llama-8B, the reductions are 96.23\%, 96.01\%, and 96.70\%. Tail (P90) latency reductions are even larger, underscoring improved user experience. On the proactive side, mean latency improvements reach 34.4\%-0.3\% for Llama-3B and 58.7\%-6.1\% for Llama-8B.

Reactive latency in \sys grows slowly with increasing proactive rates by only 19\% (3B) and 49\% (8B) from 4 to 10 proactive req/min, thanks to adaptive batching that caps batch sizes for reactive-first scheduling. Moreover, the gap between P90 and mean latency remains narrow, showing stable performance without outliers. In contrast, baselines degrade quickly under higher load: latencies inflate, tails widen, and responsiveness collapses. In baselines, reactive latencies consistently exceed proactive ones, as independent and irregular arrival patterns increase queuing delays under concurrency, and the absence of priority scheduling exacerbates delays for reactive requests. These results confirm the effectiveness of our co-scheduling policies.

\subsection{Latency Breakdown}\label{subsec:eval-latency}

\begin{figure}[t]
    \centering
    \includegraphics[width=\columnwidth]{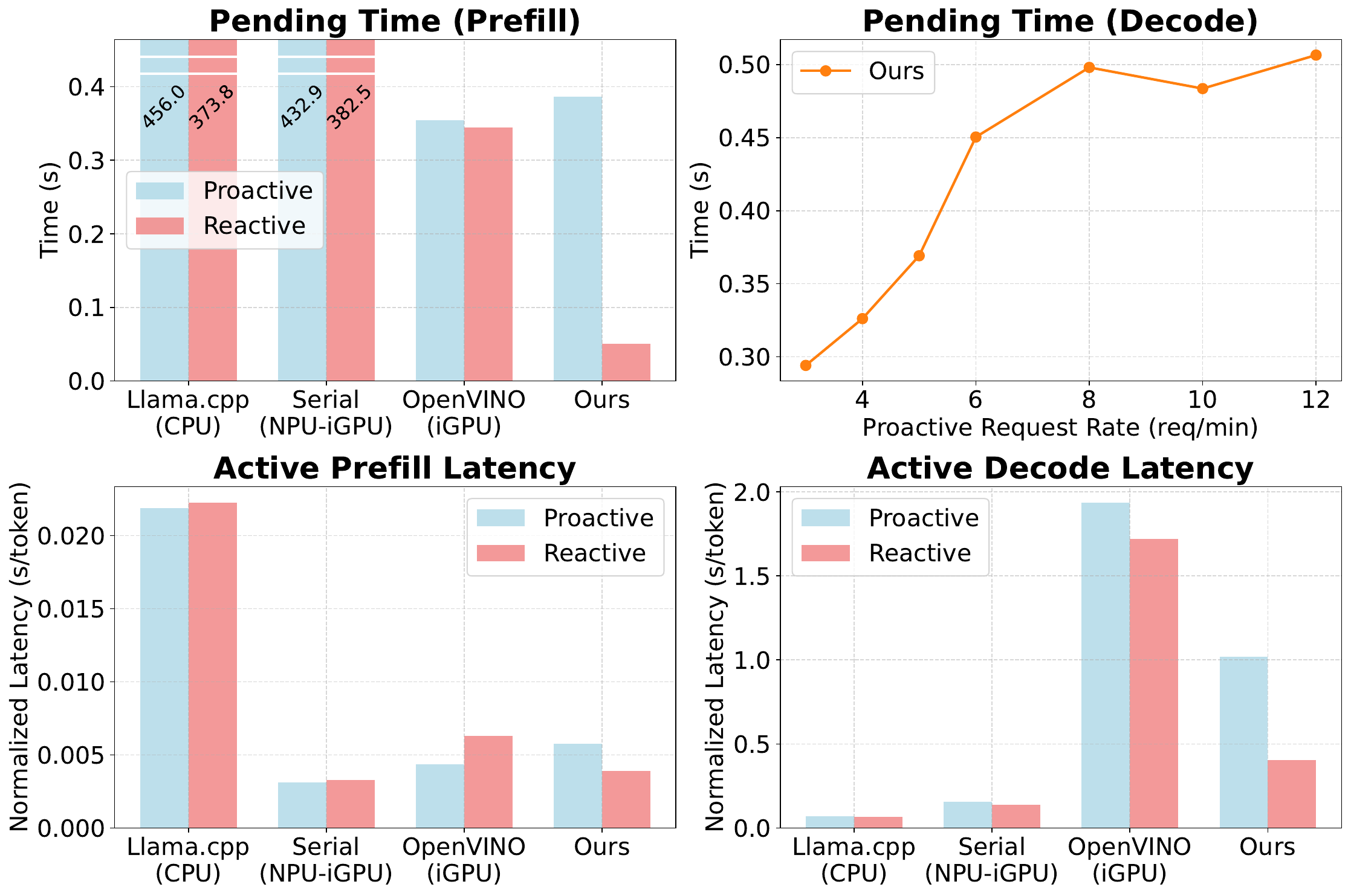}
    \caption{\textbf{Breakdown of Agentic Serving Latencies.} Most measured under mixed workloads (3 reactive and 6 proactive req/min), except the decode pending time with varying proactive rates on \sys, which is zero on baselines with serial scheduling or continuous batching.}
    \label{fig:eval-breakdown}
\end{figure}

To explain the end-to-end gains, we decompose the inference latency of each request into prefill/decode pending time, active prefill latency, and active decode latency~(\figref{fig:eval-breakdown}). The results are normalized within either proactive or reactive agentic flows. This breakdown also serves as a compact ablation.

Baselines suffer from long prefill pending times: CPU serving is delayed by slow prefill, and static NPU-iGPU inference is stalled by serial scheduling. OpenVINO (iGPU) shows significantly longer reactive pending time than \sys due to the absence of effective prioritized scheduling. In contrast, \sys keeps reactive prefill pending time to 0.048s on average, enabled by kernel-level sub-100 ms preemption. Decode pending appears only in \sys due to its unique decoupled scheduling, and grows with proactive load until saturated by the maximum reactive-first batch size.

Active prefill speeds of \sys and OpenVINO (iGPU) are similar, but \sys accelerates reactive prefill via NPU-iGPU parallelism. Serial~(iGPU-NPU) has uniformly better performance than \sys with optimal workload partition for both reactive and proactive inputs, despite its limitation to static inference. Decode shows different trends: CPU benefits from multithreading and large caches, while Serial~(NPU-iGPU) reflects fast single-batch iGPU decoding. iGPU decode latencies increase due to co-batching with long prefills, but \sys reduces reactive decode interference through reactive-first batching and fine-grained iGPU arbitration. This demonstrates why reactive requests consistently outperform proactive ones under our design.

\subsection{Overhead Analysis}\label{subsec:eval-overhead}

\begin{figure}[t]
    \centering
    \includegraphics[width=\columnwidth]{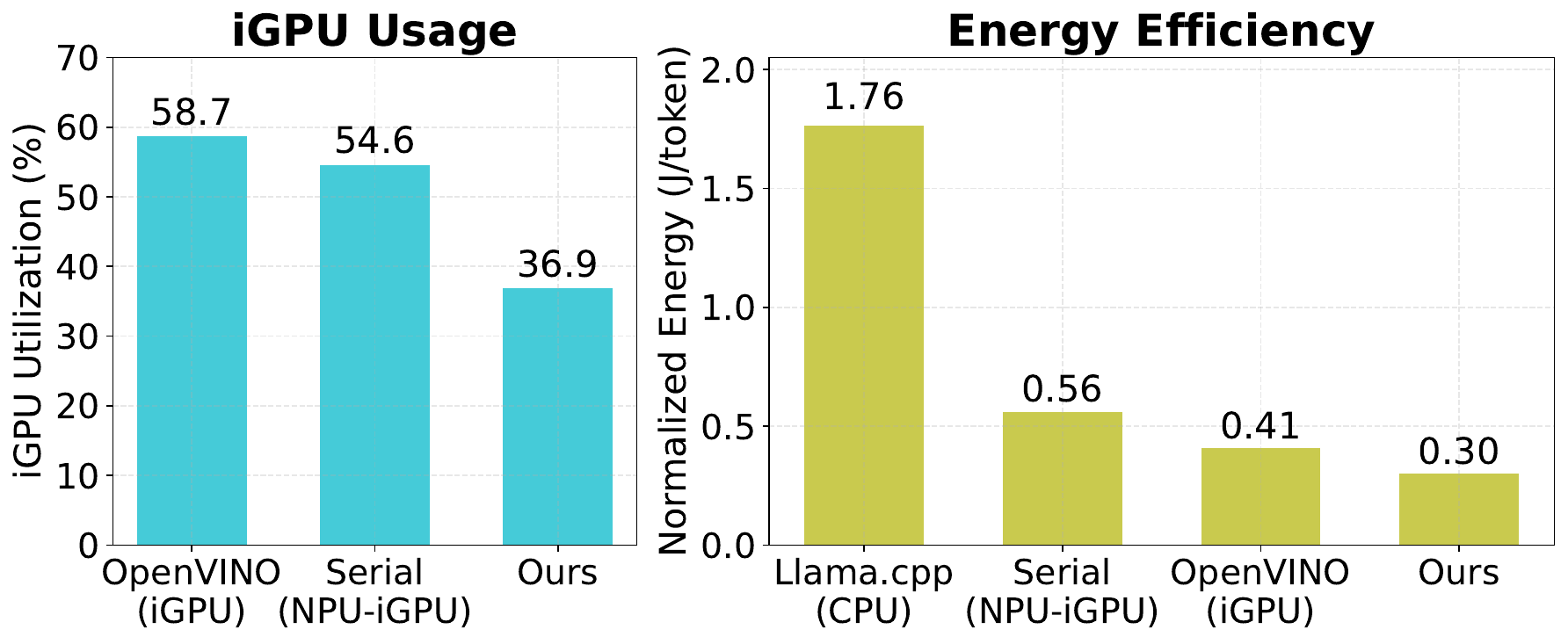}
    \caption{\textbf{Overhead Analysis.} iGPU usage and energy efficiency under the same mixed workloads as \figref{fig:eval-breakdown}.}
    \label{fig:eval-overhead}
\end{figure}

\minisection{iGPU Utilization.}
We measure iGPU utilization across inference stages on the same mini PC running Windows 11, using the system resource monitor. Pure-iGPU prefill can strike 100\% utilization, while decode averages 46\%. The higher-than-expected decode utilization (relative to FLOPS) arises from memory traffic and synchronization overheads. We account only for active prefill and decode time, excluding duplicated computation from batching. As shown in \figref{fig:eval-overhead}, \sys reduces overall iGPU utilization by 32.5\% and 37.1\% compared with Serial~(NPU-iGPU) and OpenVINO (iGPU). Although NPU offloading reduces iGPU occupation in NPU-iGPU baseline, the serial execution without decode batching increases accumulative iGPU usage over all requests.

\minisection{Energy Efficiency.}
Using Intel VTune~\cite{vtune}, we measure that NPU power remains stable around 10W for chunked prompt length. iGPU power grows from 25W (decode) to 31W (prefill), with little sensitivity to batch size ($\leqslant$32). CPU workloads consume 12W during single-threaded NPU kernel compilation and up to 58W under multi-threaded inference. \figref{fig:eval-overhead} compares per-token energy: Llama.cpp (CPU) is the least efficient, followed by Serial~(NPU-iGPU) with deficient single-batch iGPU decode. \sys achieves 26.8\% lower energy consumption than OpenVINO (iGPU) by offloading prefill to the NPU and employing efficient batching.

\section{Related Work}  

\minisection{Stateful LLM Serving.}  
Recent systems address stateful and agentic LLM serving with caching or workflow optimizations. InferCept~\cite{infercept} intercepts intermediate states to avoid recomputation, while Parrot~\cite{parrot} and Ayo~\cite{ayo} expose application-level dataflows or primitive graphs for coordinated execution. Autellix~\cite{autellix} and SGLang~\cite{sglang} generalize LLM agents into program-like structures with specialized runtimes. These systems target multi-tenant cloud settings with abundant resources, whereas our focus is on single-device agents operating under tight SoC constraints.  

\minisection{On-Device LLM Inference.}  
On-device inference efforts have explored heterogeneous accelerator use and architectural specialization. HeteroInfer~\cite{heterollm, heteroinfer} and LLM.npu~\cite{mllm-npu} exploit NPU-GPU (or CPU) parallelism, while PowerInfer-2~\cite{powerinfer-2} adopts neuron-cluster decomposition for polymorphic execution on smartphones. These works primarily optimize isolated inference latency relying on accuracy-reducing 4-bit quantization, and lack support for stateful workloads. Orthogonal researches include 1) hybrid execution among server-grade accelerators~\cite{moon2025hybe, kim2025lia, chen2025ktransformers}, which has different assumptions from on-device scenarios, and 2) co-design with novel memory architecture~\cite{li2025h2, heo2024neupims, facil} or in-storage computing~\cite{lincoln}, which requires specialized hardware and limits deployability on personal devices.  

\minisection{Preemptive Scheduling of DNN Workloads.}  
Preemption mechanisms improve responsiveness under concurrent deep learning workloads. 
PipeSwitch~\cite{bai2020pipeswitch} pipelines GPU context switching, PREMA~\cite{choi2020prema} predicts phases for NPU preemption, REEF~\cite{han2022microsecond} achieves microsecond-scale preemption via GPU reset. Pantheon~\cite{han2024pantheon} enables fine-grained preemption through sliced execution and early exits. 
XSched~\cite{shen2025xsched} proposes unified abstractions for XPU-based preemption. 
While effective for traditional DNN tasks on high-end GPUs or customized NPUs, these techniques are not applicable to the unique challenges of scheduling reactive and proactive agentic LLM flows on hetero-SoC.

\section{Conclusion}

The rise of personal LLM agents demands efficient on-device execution, yet the interplay of proactive and reactive workloads remains challenging for heterogeneous SoCs. \sys effectively bridges the three fundamental gaps hindering current ecosystems: it reconciles the flexibility-efficiency trade-off via a Heterogeneous Execution Graph that enables affinity-guided, elastic kernel mapping; it mitigates shared-memory contention through stage-decoupled scheduling that orchestrates NPU-iGPU memory access patterns; and it fills the void of flow-aware runtime abstractions by introducing prioritized coordination mechanisms, including fine-grained preemption and slack-aware piggybacking. By strictly aligning heterogeneous acceleration with agentic flow characteristics, our work paves the way for future edge platforms to serve diverse agentic LLM flows under tight resource constraints.

\newpage
\bibliographystyle{plain}
\bibliography{main}

\newpage
\appendix
\section{Analytical Modeling of Kernel Metrics}\label{sec:perf-model}

In \sys, the HEG models LLM execution as a parametric, accelerator-aware multigraph with static NPU kernels and dynamic iGPU kernels. Each HEG node carries profiling-guided annotations (latency and bandwidth hints, as a function of batch/sequence), enabling the \sys coordinator to bind elastic kernels to accelerators based on computation affinity, runtime elasticity, and memory awareness. To support these annotations, we model a hetero-SoC accelerator kernel by its total floating-point work $W$ [FLOPs], total data movement $Q$ [Bytes], and arithmetic intensity $I \equiv W/Q$ [FLOPs/Byte]. Let the target accelerator (NPU or iGPU) provide a peak compute throughput $P_{\text{pk}}$ [FLOP/s] and a peak off-chip memory bandwidth $B_{\text{pk}}$ [B/s]. The classical roofline bound on achievable performance is
$$P(I) \le \min(P_{\text{pk}}, B_{\text{pk}} I). \quad (1)$$
To capture non-idealities in operator mapping, we calibrate effective ceilings using two empirical anchor measurements that bracket the operating regimes: (i) a most memory-bound case at intensity $I_{\text{m}}$ with measured bandwidth utilization $u_{\text{m}} \in (0, 1]$ (achieved bandwidth divided by $B_{\text{pk}}$), and (ii) a most compute-bound case at intensity $I_{\text{c}}$ with measured bandwidth utilization $u_{\text{c}} \in (0, 1]$. From these we define
$$B_{\text{eff}} \equiv u_{\text{m}} B_{\text{pk}}, \quad (2)$$
$$P_{\text{eff}} \equiv \min(P_{\text{pk}}, u_{\text{c}} B_{\text{pk}} I_{\text{c}}), \quad (3)$$
which are the calibrated memory and compute ceilings, respectively. Equation (2) reflects that in the memory-bound anchor the achieved bandwidth is $u_{\text{m}} B_{\text{pk}}$. Equation (3) follows because, in the compute-bound regime, achieved bandwidth equals $P/I$; hence $u_{\text{c}} = (P/I)/B_{\text{pk}}$ implies $P_{\text{eff}} \approx u_{\text{c}} B_{\text{pk}} I_{\text{c}}$, clipped by $P_{\text{pk}}$.
Define the knee (transition) intensity
$$I_{\star} \equiv \frac{P_{\text{eff}}}{B_{\text{eff}}} = \frac{P_{\text{eff}}}{u_{\text{m}} B_{\text{pk}}} = \frac{u_{\text{c}}}{u_{\text{m}}} I_{\text{c}} \quad \text{if } P_{\text{eff}} = u_{\text{c}} B_{\text{pk}} I_{\text{c}}. \quad (4)$$

\subsection*{Latency}

The performance \( P(I) \) and bandwidth \( b(I) \) metrics, formally defined in the subsequent section, are used here to model latency. Let $T(I)$ denote the runtime (latency). Under the roofline assumption that runtime is governed by the slower of compute or memory,

$$T(I) = \max\left( \frac{W}{P_{\text{eff}}}, \frac{Q}{B_{\text{eff}}} \right) = \frac{W}{P(I)} = \frac{Q}{b(I)}. \quad (5)$$

A convenient unified expression is

$$T(I) = \frac{Q}{B_{\text{pk}} u(I)} = \frac{W}{B_{\text{pk}} u(I) I}. \quad (6)$$

Then $b(I) = B_{\text{pk}} u(I)$ and $P(I) = I b(I) = B_{\text{pk}} u(I) I$.

Piecewise forms. Explicitly,
$$T(I) = \begin{cases} \frac{Q}{u_{\text{m}} B_{\text{pk}}} & I \le I_{\star} \quad (\text{memory-bound}) \\ \frac{W}{P_{\text{eff}}} & I \ge I_{\star} \quad (\text{compute-bound}) \end{cases} \quad (7)$$
$$u(I) = \begin{cases} u_{\text{m}} & I \le I_{\star} \\ \frac{u_{\text{c}} I_{\text{c}}}{I} & I \ge I_{\star} \end{cases} \quad (8)$$

\subsection*{Performance and Bandwidth}
Achieved performance and bandwidth are derived as:
$$P(I) = \min(P_{\text{eff}}, B_{\text{eff}} I), \quad (9)$$
$$b(I) = \min(B_{\text{eff}}, P_{\text{eff}}/I). \quad (10)$$
Equivalently, using only the two anchor utilizations ($u_{\text{m}}$, $u_{\text{c}}$) and $I_{\text{c}}$, the achieved bandwidth utilization has a closed-form expression:
$$u(I) \equiv \frac{b(I)}{B_{\text{pk}}} = \min\left( u_{\text{m}}, \frac{u_{\text{c}} I_{\text{c}}}{I} \right). \quad (11)$$
Remarks and assumptions. (i) This model assumes a single dominant off-chip memory roof. If multiple memory levels matter, take $B_{\text{eff}}$ as the active ceiling for the kernel at the given $I$. (ii) The compute-bound anchor ($I_{\text{c}}$, $u_{\text{c}}$) should be sufficiently high intensity so that $P(I_{\text{c}}) \approx P_{\text{eff}}$; otherwise $P_{\text{eff}}$ inferred by (3) will be conservative. (iii) The clipping in (3) enforces $P_{\text{eff}} \le P_{\text{pk}}$. (iv) Equations (7)--(11) provide closed-form latency and bandwidth utilization predictions at arbitrary $I$, requiring only the two measured utilizations ($u_{\text{m}}$, $u_{\text{c}}$), their associated intensity $I_{\text{c}}$, and the problem size through ($W$, $Q$) or $I = W/Q$. These predictions guide runtime elasticity in prefill ops, such as partitioning into chunked NPU kernels and dynamic iGPU kernels, depending on request priority, accelerator load, and bandwidth pressure.

\end{document}